\documentclass{article}

\usepackage{graphicx}
\usepackage{float}
\usepackage{epsfig}
\usepackage{amsmath,latexsym}
\begin{document}

\title{\bf Thin accretion disk signatures of scalarized black holes in Einstein-scalar-Gauss-Bonnet gravity}
\author{{Mohaddese Heydari-Fard$^{1}$\thanks{Electronic address: m\_heydarifard@sbu.ac.ir} and Hamid Reza Sepangi$^{1}$\thanks{Electronic address: hr-sepangi@sbu.ac.ir}}\\ {\small \emph{$^{1}$ Department of Physics, Shahid Beheshti University, G. C., Evin, Tehran, Iran}}}

\maketitle

\begin{abstract}
Einstein-scalar-Gauss-Bonnet gravity has recently been known to exhibit spontaneous scalarization. In the presence of the Gauss-Bonnet term the no-hair theorem can be evaded and novel black hole solutions with non-trivial scalar fields have been found besides the general relativistic solutions. In this paper, we aim to investigate the possibility of observationally testing Einstein-scalar-Gauss-Bonnet gravity using thin accretion disk properties around such scalarized black holes. Using the Novikov-Thorne model, we numerically calculate the electromagnetic flux, temperature distribution, emission spectrum, innermost stable circular orbits and energy conversion efficiency of accretion disks around such black holes and compare the results with the standard general relativistic Schwarzschild solution. We find that the accretion disks around scalarized black holes are hotter and more luminous than in general relativity.
\vspace{5mm}
\vspace{1mm}
\end{abstract}
\section{Introduction}
Over the last century, general relativity (GR) has been exceedingly successful as a theory in describing gravitational phenomena. The discovery of gravitational waves resulting from a binary black hole merger \cite{wave}, and the first released image of a black hole shadow by the Event Horizon Telescope \cite{shadow}, are some of the recent success stories. However, there are some theoretical and observational issues which suggest modification of  GR. For example, it is found that the one-loop renormalization of GR needs to supplement Einstein-Hilbert action with second-order curvature invariants \cite{stell}. Also, some cosmological problems such as the inflationary era in the early universe and recent accelerated expansion of the universe, motivate us to seek alternative ideas beyond GR.

Among generalizations of GR are the scalar-tensor theories where a scalar field is non-minimally coupled to curvature and interfere in the generation of gravitational interaction \cite{scalar}. It is worth mentioning that theories such as Kaluza-Klein and string which attempt to unify fundamental forces naturally result in scalar-tensor generalization of GR.
The study of compact objects such as black holes and neutron stars in modified theories of gravity is an important subject and the question arises as to whether a scalar field has an imprint on black holes to distinguish them from those in GR. An interesting mechanism which has attracted renewed attention in recent years and leads to different black hole solutions from GR is the spontaneous scalarization. This phenomena was first proposed for neutron stars in the context of standard scalar-tensor theory in 1990 \cite{Damour}, where the coupling between matter and scalar field leads to an effective mass for the scalar field. If such an effective mass becomes tachyonic, the GR solution will be unstable against scalar perturbations. Therefore,  in such a scenario perturbations grow spontaneously and finally settle to a stable configuration with a non-trivial profile for the scalar field. The generalization of this study to the case of  massive scalar fields or  vector fields has been considered in \cite{Fethi1}--\cite{Cardoso}. Black holes cannot be scalarized in this way, because the Ricci scalar vanishes in vacuum. However as shown in \cite{Pani}, when a black hole is surrounded by matter, a similar mechanism can also occur. In a similar fashion, Einstein-Maxwell-scalar model was studied in \cite{Font}--\cite{Astefanesei1} where the scalar field is non-minimally coupled to electromagnetic field for which the spontaneous scalarization of charged black holes may occur. In addition to the previous cases, a recent new type of scalarization, the so-called curvature-induced scalarization has been introduced \cite{Doneva1}--\cite{Kanti1}. In Einstein-scalar-Gauss-Bonnet (EsGB) theory containing a non-minimal coupling between the scalar field and Gauss-Bonnet (GB) invariant and under specific conditions on the coupling function, hairy black hole solutions as well as GR solutions may exist. In these models the tachyonic instability of GR solutions can be triggered due to the scalar-GB coupling and new black hole solutions with non-trivial scalar fields may form. The stability of these solutions depends on the functional form of the scalar coupling and has been studied in \cite{Doneva2}--\cite{Ikeda}. The curvature-induced scalarization of charged black holes, rotating black holes, black holes with a massive scalar field and black holes with  cosmological constant has been studied in \cite{charge}--\cite{lambda2}, respectively. The study of scalarization of neutron stars and charged wormholes in EsGB gravity has been carried out in \cite{star} and \cite{wormhole}. Also, a complete analysis of higher dimensional generalizations of spontaneous scalarization models was recently studied in \cite{Astefanesei2}.

It is well known that in a binary system the accretion process can take place around compact objects (a black hole or a neutron star) where strong gravitational effects are important \cite{x1}--\cite{x2}. In this process the hot gas falls into the gravitational potential of the compact object and releases the energy in the form of heat and radiation. The properties of the emission spectra from the disk depend on the geodesic motion of particles and may also be associated with the structure of the central object. Therefore the study of disk properties can be used to test gravity in these extreme regions and explore possible deviations from GR and generalized theories of gravity. In this regard thin accretion disks have been investigated in $f(R)$ modified gravity, Horava-Lifshitz gravity, scalar-vector-tensor gravity, brane-world scenarios and Einstein-Maxwell-dilaton gravity \cite{fR}--\cite{dilaton2}. Thin accretion disks around gravastars, boson and fermion stars, naked singularities and exotic matter such as wormholes have been studied in \cite{grava}--\cite{wormhole2}. It has also been shown that the continuum-fitting method and analysis of relativistic iron line profiles are available techniques which can be used to distinguish  different astrophysical objects through their accretion disks \cite{m1}--\cite{m3}. Moreover, the effects on relevant accretion properties for black hole solutions with non-trivial scalar fields have also been studied. For instance, In Einstein-dilaton-Gauss-Bonnet gravity it is pointed out that depending on the values of the dilatonic charge and  mass of the solution, the ISCO location can be different relative to that of GR \cite{pani1}. Also, for scalarized neutron stars in scalar-tensor theories the effects of scalarization on the epicyclic frequencies,  ISCO location and  shape of the iron line have been investigated in \cite{s1} and \cite{s2}, respectively. In the present paper we study the accretion process in thin disks around scalarized black holes in EsGB gravity and investigate the effects of scalarization on their properties.

The paper is structured as follows. In Section 2, we present the EsGB theory of gravity and introduce the spontaneous scalarization mechanism in this theory. A brief review of geodesic motion and thin accretion disk model in a general static and spherically symmetric space-time is given in section 3. Our results are presented in section 4, where the disk properties around a scalarized black hole in EsGB gravity are obtained and the effects of the scalar hair are discussed. The paper ends with drawing conclusions.

\section{Einstein-scalar-Gauss-Bonnet gravity}
The action for EsGB gravity reads
\begin{equation}
{\cal S}=\frac{1}{16\pi}\int d^4x \sqrt{-g}\left[R-\frac{1}{2}g^{\mu \nu}\partial_{\mu}\varphi\partial_{\nu}\varphi+\lambda ^2f(\varphi){\cal G}\right],
\label{action}
\end{equation}
where $R$ is the Ricci scalar and ${\cal G}=R^2-4R_{\mu\nu}R^{\mu\nu}+R_{\mu\nu\rho\sigma}R^{\mu\nu\rho\sigma}$ is the GB invariant. The parameter $\lambda$ is the GB coupling constant and has the dimension of length. A well-known model of the theory with coupling function $f(\varphi)\sim e^{-\alpha\varphi}$ is that of the Einstein-dilaton-Gauss-Bonnet gravity which arises in the low energy limit of string theory. The first black hole solutions for this model were obtained in \cite{Mignemi}--\cite{Torii} and later on extended to slowly rotating and rapidly rotating black holes in \cite{pani1} and \cite{slow2}--\cite{rapid2}.

Now, varying action (\ref{action}) with respect to the metric $g_{\mu\nu}$ gives the gravitational field equations
\begin{equation}
G_{\mu\nu}=\frac{1}{2}\nabla_{\mu}\varphi\nabla_{\nu}\varphi-\frac{1}{4}g_{\mu\nu}\nabla ^{\rho}\varphi\nabla _{\rho}\varphi-4\lambda^2(\nabla^{\rho}\nabla^{\sigma}f(\varphi))P_{\mu\rho\nu\sigma},
\label{grav}
\end{equation}
where
\begin{equation}
P_{\mu\nu\rho\sigma}=R_{\mu\nu\rho\sigma}+g_{\mu\sigma}R_{\rho\nu}-g_{\mu\rho}R_{\sigma\nu}+g_{\nu\rho}R_{\sigma\mu}-g_{\nu\sigma}R_{\rho\mu}+
\frac{R}{2}(g_{\mu\rho}g_{\nu\sigma}-g_{\mu\sigma}g_{\nu\rho}).
\label{P}
\end{equation}
It is important to note that the field equations are of second-order as in GR. Therefore the Ostrogradski instability is avoided and the theory is free from ghosts. Variation of the action with respect to the scalar field gives the generalized Klein-Gordon equation
\begin{equation}
\Box\varphi+\lambda ^2f'(\varphi){\cal G}=0,
\label{scalar}
\end{equation}
where a prime denotes the derivative with respect to the scalar field $\varphi$. Numerical black hole solutions of the field equations have been found in \cite{Doneva1}--\cite{Kanti1}. It was shown that if the coupling function satisfies the condition $f'(\varphi_0)=0$ at some $\varphi_0=const$, the scalar equation (\ref{scalar}) is trivially satisfied and equation (\ref{grav}) reduces to that of the GR field equations. Thus, the field equations resulting from (\ref{action}) admit a GR solution with a constant scalar field $\varphi_0$. Now by considering small scalar perturbations $\delta\varphi$ around the GR solution, the linearized form of equation (\ref{scalar}) becomes
\begin{equation}
(\Box_{(0)}-m_{\rm eff}^2)\delta\varphi=0,
\label{}
\end{equation}
where $m_{\rm eff}^2=-\lambda ^2f''(\varphi){\cal G}_{(0)}$ is the effective mass squared for the scalar perturbations and subscript '$0$' refer to the background geometry. When $m_{\rm eff}^2<0$, the tachyonic instability leads to a GR solution being unstable. Therefore the perturbed scalar field grows spontaneously and new black hole solutions with a non-trivial scalar field will be produced. These scalarized solutions are characterized by the number of nodes, $n$, of the scalar field. In \cite{Doneva2}, it was shown that the solutions with $n>0$ are unstable under radial perturbations, while the first non-trivial solution with $n=0$ has a different behavior. The stability of this solution strongly depends on the choice of the scalar-GB coupling and although the solution is unstable for a quadratic coupling function it can be stable for an exponential one. Indeed, adding higher powers of the scalar field to the coupling function can stabilize the solution, \cite{Silva2}--\cite{Ikeda}.

In this work, following \cite{Doneva1}, we will use the coupling function
\begin{equation}
f(\varphi)=\frac{1}{12}[1-\exp(-6\varphi^2)],
\label{b4}
\end{equation}
which leads to non-negligible deviations from GR solutions in the strong field regime as well as satisfies the conditions for the existence of solutions with non-trivial profile of the scalar field. We will consider static and spherically symmetric solutions of the field equations and investigate the properties of the accretion disks around them.

\section{Thin accretion disk model}
In this section we first briefly review the equatorial circular orbits in a general static, spherically symmetric space-time and then derive the basic equations that describe the electromagnetic properties of a thin accretion disk \cite{Horava}.

The geodesic motion of test particles moving around a compact object is governed by the Lagrangian
\begin{equation}
{\cal L}= \frac{1}{2}g_{\mu\nu}\dot{x^{\mu}}\dot{x^{\nu}},
\label{lagrangi}
\end{equation}
where $g_{\mu\nu}$ is the metric of the space-time and a dot denotes differentiation with respect to the affine parameter. Let us consider a static, spherically symmetric space-time with a metric in the general form
\begin{equation}
ds^2=g_{tt}dt^2+g_{rr}dr^2+g_{\theta\theta}d\theta ^2+g_{\phi\phi}d\phi ^2,
\label{100}
\end{equation}
for which we assume the components $g_{tt}, g_{rr}, g_{\theta\theta}$ and $g_{\phi\phi}$ only depend on the radial coordinate $r$. Using the Euler-Lagrange equations, in the equatorial plane $\theta=\frac{\pi}{2}$, one obtains
\begin{equation}
\dot{t}=-\frac{\tilde{E}}{g_{tt}},
\label{2}
\end{equation}
\begin{equation}
\dot{\phi}=\frac{\tilde{L}}{g_{\phi\phi}},
\label{3}
\end{equation}
where $\tilde{E}$ and $\tilde{L}$ are the specific energy and the specific angular momentum, respectively. Now, taking $2{\cal L}=-1$ for test particles and using equations (\ref{2}) and (\ref{3}) we find
\begin{equation}
-g_{tt}g_{rr}\dot{r}^2+V_{\rm eff}(r)=\tilde{E}^2,
\label{4}
\end{equation}
where the effective potential is defined as
\begin{equation}
V_{\rm eff}(r)=-g_{tt}\left(1+\frac{\tilde{L}^2}{g_{\phi\phi}}\right).
\label{5}
\end{equation}
For stable circular orbits using conditions $V_{\rm eff}(r)=0$ and $V_{\rm eff,r}(r)=0$, one obtains the specific energy, specific angular momentum and angular velocity $\Omega$ of particles moving in the gravitational potential of the central object as follows
\begin{equation}
{\tilde{E}}=-\frac{g_{tt}}{\sqrt{-g_{tt}-g_{\phi\phi}\Omega^2}},
\label{6}
\end{equation}
\begin{equation}
{\tilde{L}}=\frac{g_{\phi\phi}\Omega}{\sqrt{-g_{tt}-g_{\phi\phi}\Omega^2}},
\label{7}
\end{equation}
\begin{equation}
\Omega=\frac{d\phi}{dt}=\sqrt{\frac{-g_{tt,r}}{g_{\phi\phi,r}}}.
\label{8}
\end{equation}
Also, the innermost stable circular orbit, $r_{\rm isco}$, of the central object can be determined from the condition $V_{\rm eff,rr}(r)=0$ which leads to equation
\begin{equation}
{\tilde{E}^2g_{\phi\phi,rr}}+{\tilde{L}^2g_{tt,rr}}+(g_{tt}g_{\phi\phi})_{,rr}=0. \label{9}
\end{equation}

The general relativistic model of thin accretion disks has been developed by Novikov, Thorne and Page \cite {Novikov}--\cite{Thorne} which is an extension of Newtonian approach of Shakura-Sunyaev \cite{Shakura}. The model presents a geometric description of thin accretion disks for which its vertical size, $h$, is negligible compared to its horizontal size, $h\ll r$. The accretion disk is in the equatorial plan of the central object and the accreting matter moves in Keplerian orbits. Also, the disk is considered to be in a steady state, that is the accretion mass rate, $\dot{M}_0$, is assumed to be constant in time.
In the steady-state model the accreting matter is assumed to be in thermodynamical equilibrium. Therefore, the disk temperature $T(r)$ is related to energy flux, $F(r)$ through the Stefan-Boltzmann law
\begin{equation}
F(r)=\sigma_{\rm SB} T^4(r),
\label{10}
\end{equation}
where $\sigma_{\rm SB}=5.67\times10^{-5}\rm erg$ $\rm s^{-1} cm^{-2} K^{-4}$ is the Stefan-Boltzmann constant. From the conservation equations for the mass, energy and angular momentum, we obtain an expression for the radiant energy flux in terms of the specific energy, angular momentum and angular velocity of the orbiting particles. Let us first consider the energy-momentum tensor of the accreting matter in the form \cite{Novikov}--\cite{Page}
\begin{equation}
T^{\mu\nu}=\rho_0u^{\mu}u^{\nu}+2u^{(\mu}q^{\nu)}+t^{\mu\nu},\label{n1}
\end{equation}
where $u_{\mu}q^{\mu}=0$ and $u_{\mu}t^{\mu\nu}=0$. The four-velocity of the orbiting particles is denoted by $u^{\mu}$ whereas $\rho_0$, $q^{\mu}$ and $t^{\mu\nu}$ are respectively the rest mass density, energy flow vector and stress tensor of the accreting matter. From the rest-mass conservation, $\nabla_{\mu}(\rho_0u^{\mu})=0$, we find that the time averaged accretion rate, $\dot{M}_{0}$, is independent of the disk radius
\begin{equation}
\dot{M}_{0}=-2\pi\sqrt{-g}\Sigma u^r=\rm const.,\label{n2}
\end{equation}
where the time averaged surface density is defined as
\begin{equation}
\Sigma(r)=\int_{-h}^{h}\langle\rho_0\rangle dz,\label{n3}
\end{equation}
with $z$ being the cylindrical coordinate. Using the conservation laws for energy, $\nabla_{\mu}E^{\mu}=0$, and angular momentum, $\nabla_{\mu}J^{\mu}=0$, we find
\begin{equation}
[\dot{M}_{0}\tilde{E}-2\pi\sqrt{-g}\Omega W^{r}_{\phi}]_{,r}=4\pi rF(r)\tilde{E},\label{n4}
\end{equation}
and
\begin{equation}
[\dot{M}_{0}\tilde{L}-2\pi\sqrt{-g}W^{r}_{\phi}]_{,r}=4\pi rF(r)\tilde{L},\label{n5}
\end{equation}
where the averaged torque $W^{r}_{\phi}$ is given by
\begin{equation}
W^{r}_{\phi}=\int_{-h}^{h}\langle t^{r}_{\phi}\rangle dz,\label{n6}
\end{equation}
and $\langle t^{r}_{\phi}\rangle$ is the $(\phi,r)$ component of the stress tensor, averaged over the time scale $\Delta t$ and over angle $\Delta\phi=2\pi$. Now, by employing the
energy-angular momentum relation $\tilde{E}_{,r}=\Omega\tilde{L}_{,r}$ and eliminating $W^{r}_{\phi}$ from equations (\ref{n4}) and (\ref{n5}), the time averaged energy flux emitted from the surface of an accretion disk is given by
\begin{equation}
F(r)=-\frac{\dot{M}_{0}\Omega_{,r}}{4\pi\sqrt{-g}\left(\tilde{E}-\Omega \tilde{L}\right)^2}\int^r_{r_{\rm isco}}\left(\tilde{E}-\Omega \tilde{L}\right) \tilde{L}_{,r}dr.
\label{11}
\end{equation}
The observed luminosity $L(\nu)$ has a red-shifted black body spectrum
\begin{equation}
L(\upsilon)=4\pi d^2 I(\nu)=\frac{8\pi h \cos\gamma}{c^2}\int_{r_{\rm in}}^{r_{\rm out}}\int_0^{2\pi}\frac{ \nu_e^3 r dr d\phi}{\exp{[\frac{h\nu_e}{k_{\rm B} T}]}-1},\label{12}
\end{equation}
where $h$ and $k_{\rm B}$ are the Planck and Boltzmann constants, respectively. Also, $\gamma$ is the disk inclination angle and $r_{\rm in}$ and $r_{\rm out}$ are inner and outer radii of the edge of the disk. The emitted frequency is given by $\nu_{e}=\nu (1+z)$, where the redshift factor $z$ can be written as
\begin{equation}
1+z=\frac{1+\Omega r\sin\phi\sin\gamma}{\sqrt{-g_{tt}-\Omega ^2g_{\phi\phi}}}.\label{120}
\end{equation}

Now, we define the accretion efficiency, $\epsilon$, which demonstrates the capability of the central object to convert rest mess into radiation. This quantity is defined as the ratio of the rate of the energy of photons escaping from the disk surface to infinity, and the rate at which mass-energy is transported to the black hole \cite{Novikov}--\cite{Page}. Assuming all emitted photons can escape to infinity, the radiative efficiency in terms of the specific energy measured at the ISCO radius is given by
\begin{equation}
\epsilon=1-\tilde{E}_{\rm isco}.\label{13}
\end{equation}

\section{Numerical results}
In the following, we will consider static, spherically symmetric black hole solutions in EsGB gravity with an ansatz for the metric as follows
\begin{equation}
ds^2=-e^{2\Phi(r)}dt^2+e^{2\Lambda(r)}dr^2+r^2(d\theta^2+\sin^2\theta d\phi^2).
\label{b1}
\end{equation}
Then, Einstein's equations (\ref{grav}) read
\begin{equation}
\frac{2}{r}\left[1+\frac{2}{r}(1-3e^{-2\Lambda})\Psi_r\right]\frac{d\Lambda}{dr}+\frac{(e^{2\Lambda}-1)}{r^2}
-\frac{4}{r^2}(1-e^{-2\Lambda})\frac{d\Psi_r}{dr}-\left(\frac{d\varphi}{dr}\right)^2=0,
\label{c1}
\end{equation}

\begin{equation}
\frac{2}{r}\left[1+\frac{2}{r}(1-3e^{-2\Lambda})\Psi_r\right]\frac{d\Phi}{dr}-\frac{(e^{2\Lambda}-1)}{r^2}-\left(\frac{d\varphi}{dr}\right)^2=0,
\label{c2}
\end{equation}

\begin{equation}
\frac{d^2\Phi}{dr^2}+\left(\frac{d\Phi}{dr}+\frac{1}{r}\right)\left(\frac{d\Phi}{dr}-\frac{d\Lambda}{dr}\right)+\frac{4e^{-2\Lambda}}{r}
\left[3\frac{d\Phi}{dr}\frac{d\Lambda}{dr}-\frac{d^2\Phi}{dr^2}-\left(\frac{d\Phi}{dr}\right)^2\right]\Psi_r
-\frac{4e^{-2\Lambda}}{r}\frac{d\Phi}{dr}\frac{d\Psi_r}{dr}+\left(\frac{d\varphi}{dr}\right)^2=0,
\label{c3}
\end{equation}
where $\Psi_r=\lambda ^ 2f'(\varphi)\frac{d\varphi}{dr}$ and the scalar field equation (\ref{scalar}) is given by
\begin{equation}
\frac{d^2\varphi}{dr^2}+\left(\frac{d\Phi}{dr}-\frac{d\Lambda}{dr}+\frac{2}{r}\right)\frac{d\varphi}{dr}-\frac{2\lambda ^2}{r^2}f'(\varphi)
\left[(1-e^{-2\Lambda})\left[\frac{d^2\Phi}{dr^2}+\frac{d\Phi}{dr}\left(\frac{d\Phi}{dr}-\frac{d\Lambda}{dr}\right)\right]+2e^{-2\Lambda}\frac{d\Phi}{dr}\frac{d\Lambda}{dr}\right]=0
.\label{c5}
\end{equation}
To obtain the black hole solutions with non-trivial scalar hair we have to solve  the above field equations numerically by employing the shooting method. In our numerical analysis, for simplicity, we set $r_H=1$ and integrate the field equations (\ref{c1})-(\ref{c5}) from $r=r_{H}+\epsilon$ with $\epsilon=10^{-5}$, to $r\rightarrow\infty$. Here, we only state the results and refer the interested reader to \cite{Doneva1} for the details of calculations. The requirement for the regularity of the scalar field and its first and second derivatives on the horizon is given by
\begin{equation}
r_H^4>24\lambda^4f'^2(\varphi_H),
\label{b2}
\end{equation}
where $\varphi_H$ is the value of the scalar field on the horizon, $r_H$. The mass of the black hole $M$ and the dilaton charge $D$ can be obtained from asymptotic behavior of functions $\Lambda$, $\Phi$ and $\varphi$ which are given by
\begin{equation}
\Lambda\approx\frac{M}{r}+{\cal{O}}(1/r^2),\quad \Phi\approx-\frac{M}{r}+{\cal{O}}(1/r^2), \quad \varphi\approx\frac{D}{r}+{\cal{O}}(1/r^2).
\label{b3}
\end{equation}
Also as pointed out in \cite{Doneva1}, the dilaton charge is a secondary hair; it is not an independent quantity but instead it depends on the black hole mass.

Now, one is ready to study properties of thin accretion disks around scalarized black holes in EsGB gravity and compare the results with that of the Schwarzschils in GR. As was mentioned before, these new GB black hole solutions with curvature induced scalarization were first obtained numerically in \cite{Doneva1}. Since the non-trivial solutions with a scalar field that has one or more nodes are unstable \cite{Doneva2}, in our study we will consider only the first non-trivial solutions with the scalar field with zero nodes.

The profile of the scalar field as a function of radial coordinate is shown for some values of the coupling constant $\lambda$ in the left panel of figure 1. Also, the behavior of the effective potential for these scalarized black holes is shown in the right panel of figure 1 and for comparison we have plotted the corresponding results for the Schwarzschild black hole. As is clear, the effective potential is larger in EsGB gravity than in GR and with increasing $\lambda$, the deviation from GR also increases.
\begin{figure}[H]
\centering
\includegraphics[width=3.0in]{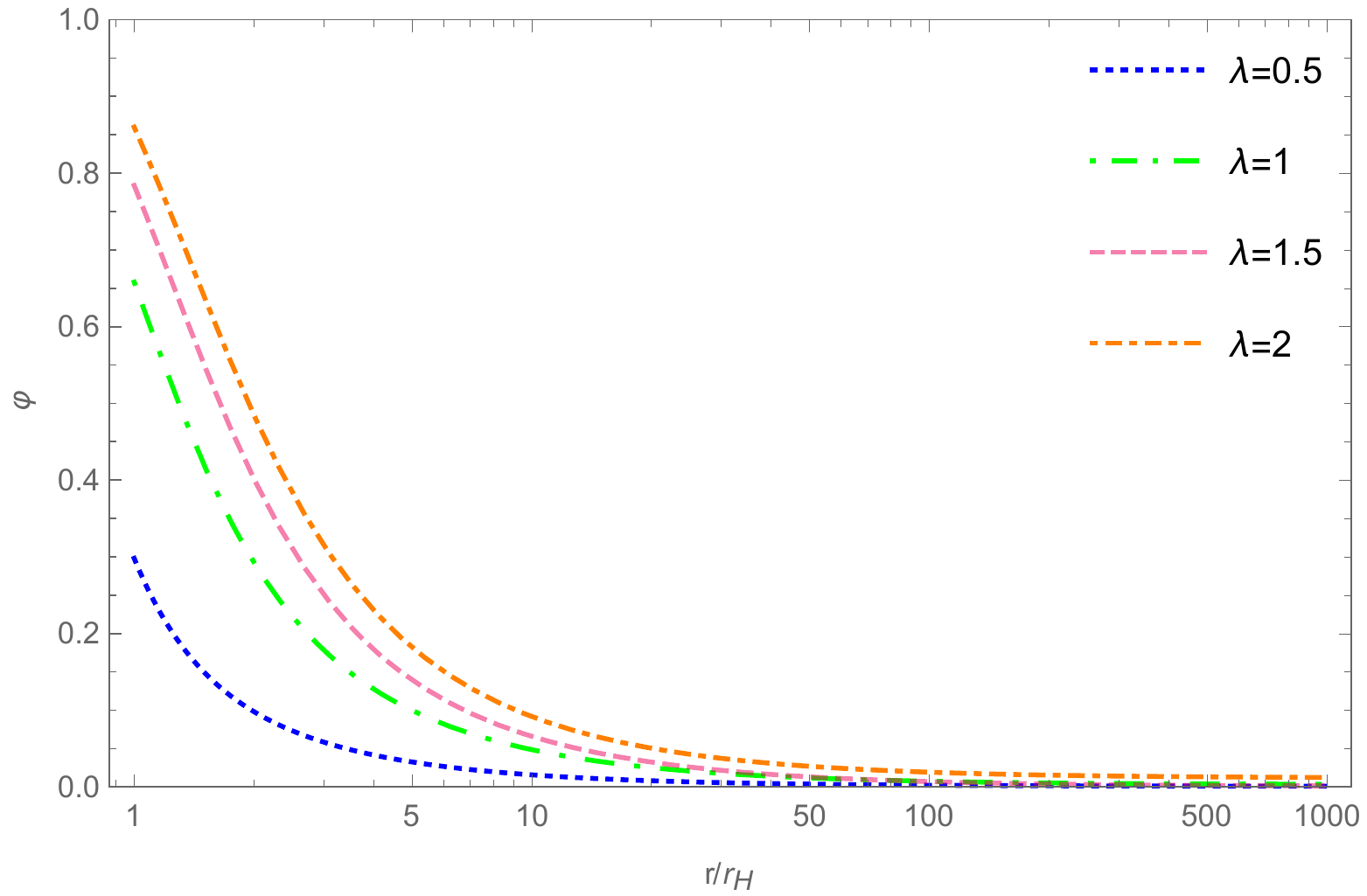}
\includegraphics[width=3.0in]{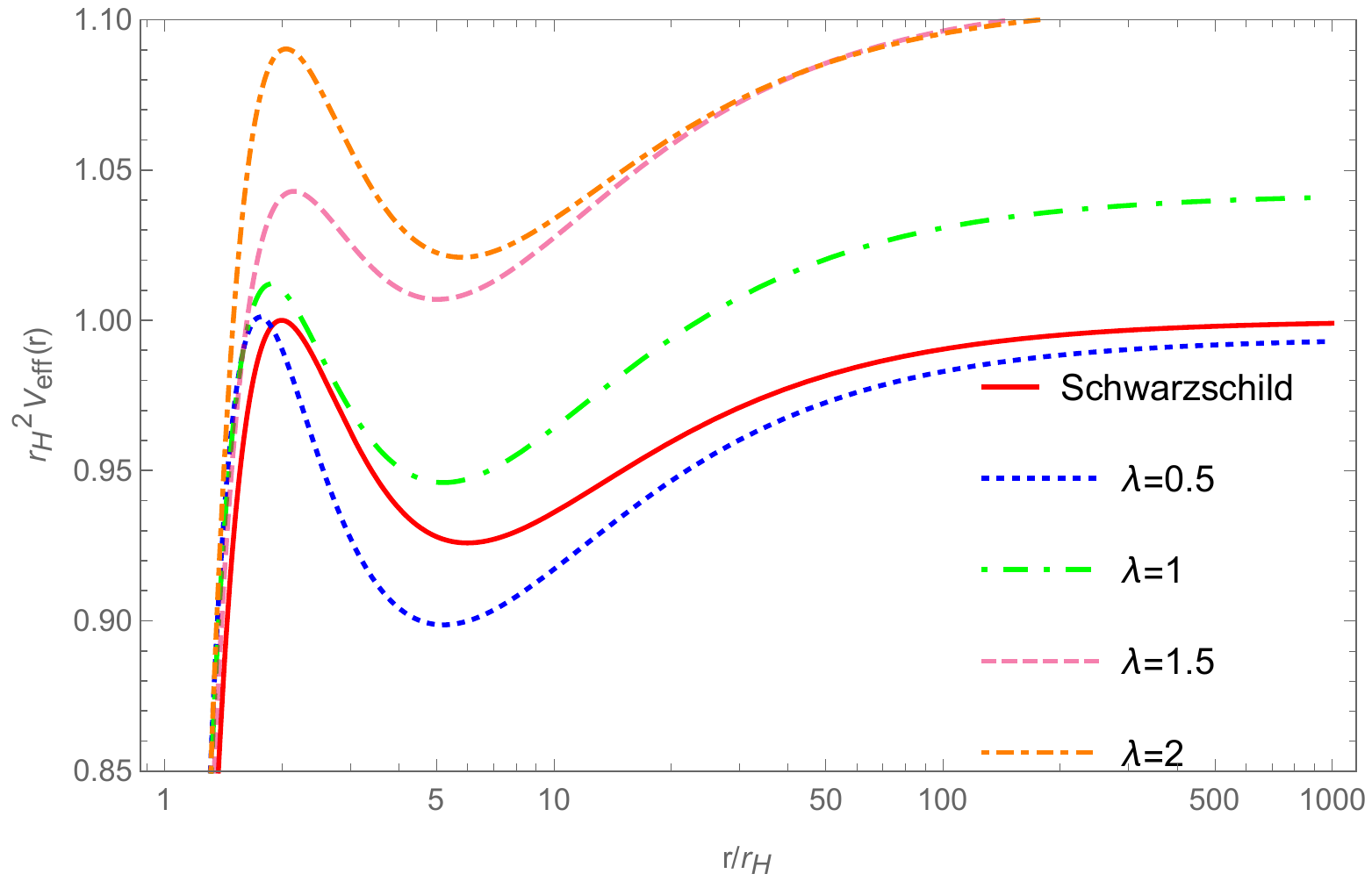}
\caption{The scalar field, left panel and the effective potential multiplied by $r_{H}^{2}$, right panel, as  functions of the normalized radial coordinate $r/r_{H}$ for some black hole solutions with different values of $\lambda$. In the right panel the solid curve corresponds to the Schwarzschild black hole.}
\label{potential}
\end{figure}

\begin{figure}[H]
\centering
\includegraphics[width=3.0in]{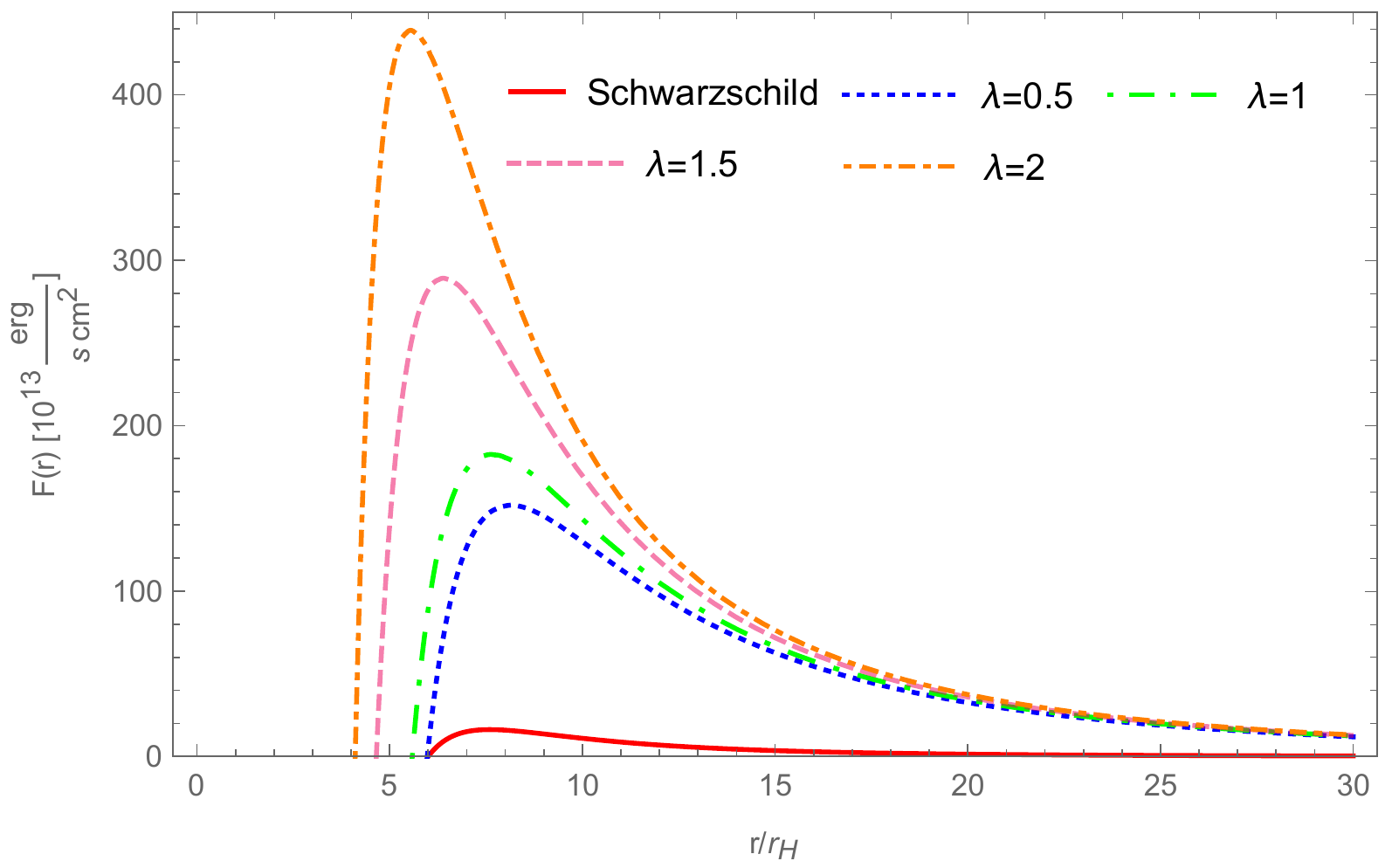}
\includegraphics[width=3.0in]{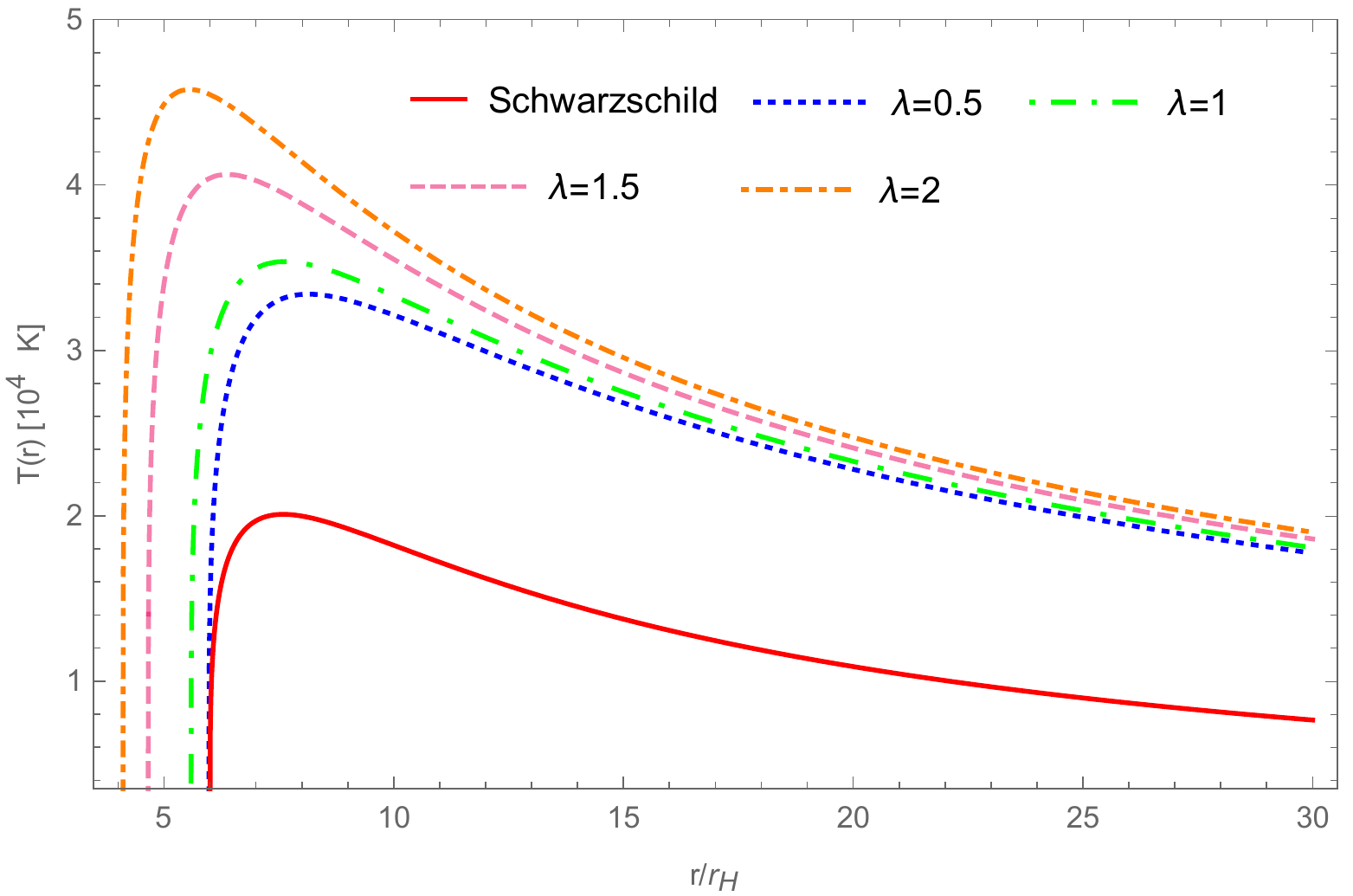}
\caption{The energy flux $F(r)$ of a disk around a static scalarized black hole with mass accretion rate $\dot{M}=2\times10^{-6}M_{\odot}yr^{-1}$ for different values of $\lambda$, left panel, and the disk temperature $T(r)$, right panel, as functions of the normalized radial coordinate $r/r_{H}$. In each panel the solid curve corresponds to the Schwarzschild black hole.}
\label{flux}
\end{figure}

\begin{figure}[H]
\centering
\includegraphics[width=3.0in]{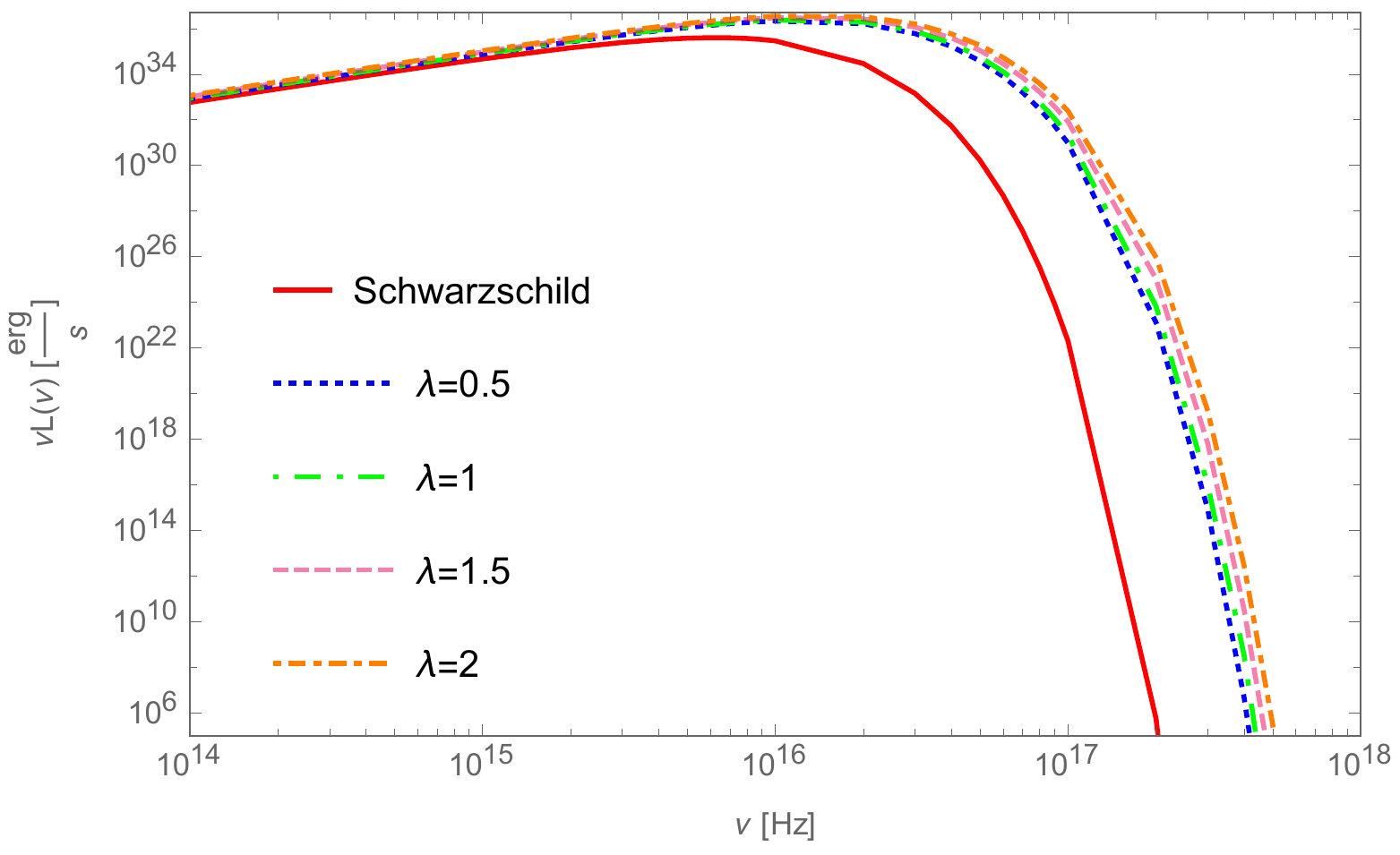}
\caption{The emission spectrum $\nu L(\nu)$ of the accretion disk around a static scalarized black hole with mass accretion rate $\dot{M}=2\times10^{-6}M_{\odot}yr^{-1}$ and inclination $\gamma=0^{\circ}$ for different values of $\lambda$, as a function of frequency $\nu$. The solid curve represents the disk spectrum for a Schwarzschild black hole.}
\label{spectra}
\end{figure}

The effects of scalarization on the energy flux over the surface of the disk for some values of $\lambda$  is shown in the left panel of figure 2. It is easy to see that for scalarized black holes the energy flux is more pronounced than in the Schwarzschild black hole (the scalar-free solution) and as the values of the GB coupling become larger the maximum of the energy flux increases. This is because with increasing GB coupling the ISCO radii of scalarized black holes become smaller and shift closer to the horizon. Therefore, for smaller ISCO radii, disk particles experience stronger gravitational fields and thus more energy is released from the gravitational potential energy. The disk temperature is plotted in the right panel of the figure and the same behavior is also observed there.

In figure 3, we display the disk spectra for scalarized black holes and for comparison we also present the emission spectra for the Schwarzschild black hole. Similar to the case of the energy flux, deviation of $L(\nu)$ for scalarized black holes from that of the Schwarzschild becomes more pronounced with increasing values of $\lambda$. Also, it is clear that with increasing GB coupling, the cut-off frequencies for scalarized black holes shift to higher frequencies.

In Table 1, we present numerical results for the dilaton charge, ISCO radius, $r_{\rm isco}$, orbital frequency, $\Omega_{\rm isco}$ and  radiative efficiency, $\epsilon$ of scalarized black holes for some values of the GB coupling. We have found that the first branch solution with non-trivial scalar field bifurcates from the Schwarzschild solution at the point $\lambda=0.42$. Also the stability analysis shows that black hole solutions with $\lambda\geq5.15$ are unstable against radial perturbations. That is why, in Table 1, we have chosen values of $\lambda$ to be in the range $0.42\leq\lambda\leq5.15$. As is clear, the ISCO radius in scalarized black holes becomes smaller and $\Omega$ becomes larger as the value of the GB coupling increases. This is due to the fact that the GB term effectively counteracts gravity, thus the instability area around scalarized black holes decreases and the ISCO radius takes smaller values. Also it is worth stressing that we have presented $\frac{r_{\rm isco}}{M}$ in Table 1 where $M$  is the physical mass defined in equation (\ref{b3}). Such a mass is measured at infinity and is interpreted as due to the contribution from the scalar field. Only in the case of the Schwarzschild black hole the physical mass is equal to the bare mass in GR and we have $\frac{r_{\rm isco}}{M}=6$. But the scalarized black holes have a larger physical mass and thus the ratio $\frac{r_{\rm isco}}{M}$ for these black holes is smaller than that for the Schwarzschild black hole. Moreover we note that  $\lambda$  can be constrained by electromagnetic and gravitational wave observations. For instance, it is found that the least massive black hole observed in x-ray binary A0620-00 and the remnant of GW170817 lead to constrains, $\lambda\leq27$ $\rm km$ and $\lambda\leq24$ $\rm km$, respectively \cite{Doneva2}. Therefore, $\lambda$ is in units of km in Table 1.

In Table 2, The maximum values of the energy flux, temperature distribution and emission spectra for scalarized black holes are presented and compared to that of the Schwarzschild's. It is easy to see that the peak values of $F_{\rm max}(r)$, $T_{\rm max}(r)$ and $\nu L(\nu)_{\rm max}$ grow with increasing  $\lambda$. Also, the cut-off frequencies for which the corresponding maxima occur, shift to higher values.

\begin{table}[H]
\centering
\caption{The $r_{\rm isco}$, orbital frequency, $\Omega_{\rm isco}$ and the efficiency for thin accretion disk around static scalarized black hole.}
\begin{tabular}{l l l l l l}
\hline
$\lambda$&$D/M$& $r_{\rm isco}/M$&$M\Omega_{\rm isco}$&$\epsilon$\\ [0.5ex]
\hline
-&0& 6.0&0.0680&0.0572\\
\\
0.5& 0.0088& 5.9761&0.0681&0.0572\\
\\
0.75& 0.2798& 5.8467&0.0719&0.0578\\
\\
1& 0.7964& 5.5872&0.0787&0.0583\\
\\
1.25& 0.9558& 5.1757&0.0855&0.0605\\
\\
1.5& 1.0935& 4.6545&0.0958&0.0632\\
\\
1.75& 1.2045& 4.2958&0.1029&0.0697\\
\\
2& 1.3029& 4.1065&0.1055&0.0885\\
\hline
\end{tabular}
\end{table}

It should be noted that this study can be extended to the case of other scalar-GB coupling functions such as even and odd polynomials and inverse polynomial functions, considered in \cite{Kanti1}, which satisfy the conditions for the existence of scalarized black holes. Moreover, we note that geodesic equations and relevant accretion properties depend only on the metric components. Therefore, one can easily apply this model to charged or spinning scalarized black holes and study thin accretion disks around them. Here we would like to mention the effects of the spin parameter on disk properties. As is well known, the accretion process around rotating black holes in GR was first studied  in \cite{Novikov}. It was found that in the presence of a rotation parameter, the ISCO radius of the disk decreases in comparison to  Schwarzschild black holes and thus the radiative efficiency of an accretion disk increases from 6\% for the Schwarzschild black hole to 42\% for a co-rotating disk around an extremal Kerr black hole \cite{Thorne}. Similarly it is expected that in the case of scalarized rotating black holes the rotation parameter also causes the ISCO radius of the disk to decrease. Indeed, as is shown in \cite{rotating1}, the relative frequencies at the ISCO for spinning scalarized black holes and Kerr black holes (namely $\frac{\Omega^{(\rm s)}}{\Omega^{(\rm GR)}}$) is greater than one. Therefore, the spinning scalarized black holes have smaller ISCO radius and thus larger energy flux in comparison with Kerr black holes. However, it is pointed out that there is a spin selection effect, so that black holes with a large spin are indistinguishable from Kerr black holes and only low spin scalarized black holes have significant deviation from GR.

Finally, we mention that the emission spectra from accretion disks have imprints of the background space-time and can be used as astrophysical probes to constrain modified theories of gravity \cite{new1}--\cite{new3}. This is called the continuum fitting method and was first proposed in \cite{new4}. In addition to rotating black holes, the continuum fitting method can also be used for non-rotating space-times to test the metric around a black hole \cite{new5}--\cite{new6}. In order to investigate whether the continuum spectrum observations can constrain the EsGB gravity, one has to theoretically  estimates  the luminosity and compare it with observation, using a minimum chi-squared, $\chi^2$, approach. The reduced $\chi^2$ is given by
\begin{equation*}
\chi_{\rm red}^2=\Sigma_{i}\frac{L_{\rm obs}(\nu_{i})-L_{\rm theo}(\nu_{i},\lambda)}{\sigma(\nu_{i})}
\end{equation*}
where $L_{\rm obs}(\nu_{i})$ denotes the observed data with possible errors $\sigma$ and $L_{\rm theo}(\nu_{i},\lambda)$ is the model estimates of the luminosity. The value of $\lambda$ that minimizes the reduced $\chi_{\rm red}^2$ is the most favored value of the GB coupling. This study is out of the scope of the present paper and will be considered elsewhere.

\begin{table}[H]
\centering
\caption{ The maximum values of the radiant energy flux $F(r)$, temperature distribution $T(r)$ and the emission spectra. The cut-off frequency is also shown in column 5. }
\begin{tabular}{l l l l l l}
\hline
$\lambda$    &$F_{\rm max}$ [\rm erg $\rm s^{-1}$ $\rm cm^{-2}$]$\times 10^{14}$&   $T_{\rm max}$ [\rm K] $\times 10^{4}$   &$\nu L(\nu)_{\rm max}$ [\rm erg $\rm s^{-1}$]$\times 10^{35}$    &$\nu_{crit}$[\rm Hz]$\times 10^{15}$\\ [0.5ex]
\hline
- & 1.6271 & 2.0084 & 3.9048 & 6.1776\\
\\
0.5 & 1.2370$\times 10$ & 3.3327 & 2.0867$\times 10$ & 9.9594\\
\\
0.75 & 1.3375$\times 10$ & 3.3780 & 2.1339$\times 10$ & 1.0002$\times 10$\\
\\
1 & 1.5553$\times 10$ & 3.5141 & 2.3150$\times 10$ & 1.0041$\times 10$\\
\\
1.25  & 1.9781$\times 10$ & 3.7409 & 2.5729$\times 10$  & 1.0079$\times 10$\\
\\
1.5 & 2.7111$\times 10$ & 4.0359 & 2.9017$\times 10$  & 1.0266$\times 10$\\
\\
1.75 & 3.5553$\times 10$ & 4.3308 &  3.1982$\times 10$ & 1.0821$\times 10$\\
\\
2 & 4.3775$\times 10$ &  4.5576  & 3.4539$\times 10$ & 1.1364$\times 10$\\
\hline
\end{tabular}
\end{table}

\section{Conclusions}
In this paper we have studied the properties of thin accretion disks around scalarized black holes in EsGB theory of gravity. We considered the coupling function  $f(\varphi)=\frac{1}{12}[1-\exp(-6\varphi^2)]$ which can lead to large deviations from GR in the strong field regime and numerically obtained physical properties of accretion disks such as the energy flux, temperature distribution and the emission spectra for some values of the GB coupling constant, using the steady-state Novikov-Thorne model, and displayed the relevant results in figures 2 and 3 for accreting scalarized black holes. We also studied the effects of non-trivial scalar fields on the ISCO radius and angular frequency $\Omega$ and showed that for scalarized black holes the ISCO radius is smaller than that of the Schwarzschild, in contrast to $\Omega$ which becomes larger, with the results summarized in Table 1. We presented characteristics of the emissivity profile of thin disks including the maximum values of the radiation energy flux, temperature distribution and emission spectra in Table 2 and showed that with increasing $\lambda$, the peaks of $F_{\rm max}(r)$, $T_{\rm max}(r)$ and $\nu L(\nu)_{\rm max}$  grow as well. Comparing the results to that in GR we found that thin accretion disks around scalarized black holes in EsGB gravity are hotter and more luminous than in GR. In this work, we focused  attention on the structure and electromagnetic properties of relativistic thin accretion disks. It would be interesting to investigate whether one can practically test the EsGB gravity and constrain the model parameter with continuum spectrum observations. We will address such problems in a future work.

\section*{Acknowledgements}
We would like to thank the anonymous referee for valuable comments.

\end{document}